\documentclass[a4paper,12pt]{article}
\usepackage{amsfonts}
\usepackage{latexsym}
\usepackage{amsmath}
\usepackage{amssymb}
\usepackage{amssymb}
\usepackage{slashed}
\usepackage{upgreek}
\usepackage{mathrsfs}
\usepackage{bbm}

\usepackage{amsthm}

\theoremstyle{remark}

\usepackage[utf8]{inputenc}
\usepackage[english]{babel}

\theoremstyle{definition}

\allowdisplaybreaks

\usepackage{relsize}
\usepackage{graphicx}
\usepackage{comment}

\renewcommand{\thefootnote}{\fnsymbol{footnote}}

\def\appendix#1{\addtocounter{section}{1}\setcounter{equation}{0}
\renewcommand{\thesection}{\Alph{section}}
\section*{Appendix \thesection\protect\indent \parbox[t]{11.15cm}{#1}}
\addcontentsline{toc}{section}{Appendix \thesection\ \ \ #1}}

\newcommand{\bea}{\begin{eqnarray}}
\newcommand{\eea}{\end{eqnarray}}

\usepackage[usenames,dvipsnames,svgnames,table]{xcolor}	
\usepackage[normalem]{ulem}				
\usepackage{microtype}
\usepackage[colorlinks, citecolor=blue, linkcolor=blue, urlcolor=Maroon, filecolor=Maroon, linktocpage=true]{hyperref} 

\usepackage{chngcntr}
\counterwithout{equation}{section}

\begin{document}

\begin{center}
\vspace*{-1.0cm}
\begin{flushright}
\end{flushright}


\vspace{2.0cm} {\Large \bf Twisted  form  hierarchies, Killing-Yano equations   and supersymmetric backgrounds } \\[.2cm]

\vskip 2cm
 G.  Papadopoulos
\\
\vskip .6cm


\begin{small}
\textit{Department of Mathematics
\\
King's College London
\\
Strand
\\
 London WC2R 2LS, UK}\\
\texttt{george.papadopoulos@kcl.ac.uk}
\end{small}
\\*[.6cm]

\end{center}

\vskip 2.5 cm

\begin{abstract}
\noindent
We show that the Killing spinor equations of all supergravity theories which may include higher order corrections on a (r,s)-signature spacetime  are associated with  twisted covariant form hierarchies. These hierarchies
are characterized by a connection on the space of forms which may not be degree preserving. As a consequence  we demonstrate that the form Killing spinor bi-linears of all supersymmetric backgrounds  satisfy a suitable generalization of  conformal Killing-Yano equation with respect to this connection.   To illustrate the general proof the twisted covariant form hierarchies of some supergravity theories in 4, 5, 6, 10 and 11 dimensions are also presented.

\end{abstract}

\newpage

\renewcommand{\thefootnote}{\arabic{footnote}}


\section{Introduction}

In the past fifteen  years there has been much progress towards the classification of solutions of supergravity theories that preserve a fraction of the supersymmetry, for a review see \cite{rev} and references within.  Two main methods have been used for this.  One is the ``bi-linears method'' which turns the Killing spinor equations (KSEs) of the theories into conditions on the form spinor bi-linears and proceeds to solve the latter \cite{pakis}. The other is the ``spinorial geometry method'' which solves the KSEs directly using spinorial techniques and the  covariance properties of the KSEs \cite{spingeom}.

The gravitino KSE of a supergravity theory is the vanishing condition of the supersymmetry variation of the gravitino evaluated at the vanishing locus of all fermionic fields of the theory. Geometrically the gravitino KSE is a parallel transport equation, ${\cal D}\epsilon=0$,  for the supersymmetry parameter\footnote{In what follows, the supersymmetry parameter $\epsilon$ is taken to be commuting spinor.}, $\epsilon$, with respect to the supercovariant  connection, ${\cal D}$,  which is constructed from the fields of the theory.  The supersymmetry variations of the remaining fermions of the theory give rise to algebraic conditions on $\epsilon$.  In the spinorial geometry method both the parallel transport equation and the algebraic ones are directly solved. In the bi-linears method, the gravitino KSE turns into a first order equation on the form Killing spinor bi-linears and  yields a set of equations that at first sight do not appear to have a direct geometric significance.

 One of the main result of this paper is to demonstrate that the conditions on the form Killing spinor bi-linears imposed by the gravitino KSE  in any supergravity theory irrespective of the spacetime signature can be organized in terms of  {\sl twisted covariant form hierarchies} (TCFHs).
The definition of a TCFH has  been given in \cite{eigen} in the context of eigenvalue estimates for certain multi-form modified Dirac operators and it is repeated below.
One of the characteristics of TCFH structure is the existence of a connection, $\nabla^{\cal F}$,  the {\sl TCFH connection},  on the space of forms which does not necessarily preserve the forms' degree. Typically there is a family TCFHs associated to the conditions imposed by the gravitino KSE on the form  spinor bi-linears.  Each TCFH in the family is distinguished by  the choice of $\nabla^{\cal F}$. For every supergravity theory, there is a maximal and a minimal choice of $\nabla^{\cal F}$ that can be made.   We shall also explain how the TCFH structures associated to a gravitino KSE depend on the choice of form Killing spinor bi-linears.  The relation described above between TCFHs and KSEs persists after including higher order corrections to supergravity theories like those for example that emerge
in the investigation of low energy effective theories for superstrings and M-theory.

A consequence of the existence of TCFHs associated to every supergravity theory is that the form Killing spinor bi-linears of all supersymmetric
backgrounds satisfy the twisted conformal Killing-Yano  equation.  This   is a suitable  generalization of the conformal Killing-Yano equation (CKY)  which is given in equation (\ref{ccy}) below, where now the characteristic connection is the TCFH connection   $\nabla^{\cal F}$ instead of $\nabla$.

Our main results will be illustrated with some examples that include the heterotic and $N=(1,0)$ $d=6$ supergravities, minimal $N=2$, $d=4$ and $N=1$, $d=5$ supergravities, and the 11-dimensional supergravity.  Some applications of the main results will be outlined in the conclusions.

It is well-known that both the Killing-Yano (KY) and CKY equations    have   applications in gravitational physics. In particular, they are used in the integrability of the geodesic systems, Hamilton-Jacobi equation,  Klein-Gordon equation and  Dirac equation  on black hole and other spacetimes, for some selected works see \cite{penrose}-\cite{lun} and for  concise reviews see \cite{revky, frolov}  and references within.
The KY and CKY equations  are also used to find the conserved charges of supersymmetric relativistic and non-relativistic
particle systems \cite{gibbons, sfetsos, hl}.

Generalizations of the KY and CKY have also been considered. In the context of supersymmetric relativistic and non-relativistic
particles, a generalization of the KY  equations has been introduced in \cite{gps, gky2} that includes skew-symmetric torsion.  A similar
generalization of CKY equation has been considered in the context of gravitational physics in \cite{warnick1, warnick2}. The relation between
KY and G-structures has been explored in \cite{gky1, gky2, santillan}.  Further generalizations of KY and CKY equations have been investigated in \cite{hl}.

\section{Twisted covariant form hierarchies}

\subsection{CKY equations and covariantly constant spinors}

Let $M$ be a n-dimensional manifold with $(r,s)$ signature equipped with a metric $g$. The CKY condition on a  k-form\footnote{We use standard conventions for the normalization of forms and for the definition of operations on the space of forms, e.g. those in \cite{rev}.} $\omega$ on $M$  is
\bea
\nabla_X\omega={1\over k+1}i_X d\omega-{1\over n-k+1} \alpha_X\wedge \delta\omega~,
\label{ccy}
\eea
where $\nabla$ is the Levi-Civita connection of  $g$, $i_X$ is the inner derivation on the space of forms with the vector field $X$, $\alpha_X(Y)=g(X,Y)$ and $\delta$ is the adjoint operation of the exterior derivative $d$. If $\delta\omega=0$, the remaining condition is the KY equation.  The condition  (\ref{ccy}) on  1-forms  implies that the associated vector field generates a conformal motion on $M$.
 A  generalization of (\ref{ccy}) is to replace the Levi-Civita connection $\nabla$ with a connection $\nabla^H=\nabla+{1\over2} H$, where $H$ is a skew-symmetric torsion, and appropriately replace $d$ and $\delta$ with $d^H$ and $\delta^H$, respectively.

It has been known for sometime that the Killing spinor bi-linears constructed from the solutions $\epsilon$ of
 the KSE\footnote{The term Killing included in KSE refers to the property of (standard) supergravity KSEs which admit an 1-form bi-linear that satisfies the Killing condition. However this property does not hold for all parallel transport equations on  spinors.  Nevertheless we shall maintain the KSEs terminology as it has become standard.}
\bea
{\cal D}_M\epsilon\equiv\nabla_M \epsilon+ \lambda\, \Gamma_M\epsilon=0~,
\eea
satisfy the CKY equation, where $\lambda$ is a complex constant. Such KSEs arise on spheres $S^n$, and  de-Sitter  dS$_n$ and  Anti-de-Sitter AdS$_n$ spaces.
To see this define the spinor bi-linears
\bea
\tau^k={1\over k!} \langle\epsilon, \Gamma_{N_1\dots N_k}\epsilon\rangle_D\, dx^{N_1}\wedge\cdots \wedge dx^{N_k}~,
\eea
where $\langle\cdot, \cdot\rangle_D$ is the Spin-invariant Dirac inner product. Although $\langle\cdot, \cdot\rangle_D$ has been used here for definiteness,
any other Spin-invariant bi-linear can be used, see e.g. appendix B of \cite{rev} for a discussion.
Next suppose that $\epsilon$ is a Killing spinor,  i.e. ${\cal D}_M\epsilon=0$, then one can show that for Lorentzian signature manifolds
\bea
\nabla_X \tau^k= (\bar \lambda-(-1)^k \lambda) i_X\tau^{k+1}+(\bar \lambda+(-1)^k \lambda) \alpha_X\wedge \tau^{k-1}~,
\label{hcon1}
\eea
while for Euclidean signature manifolds
\bea
\nabla_X \tau^k= -(\bar \lambda+(-1)^k \lambda) i_X\tau^{k+1}-(\bar \lambda-(-1)^k \lambda) \alpha_X\wedge \tau^{k-1}~.
\label{hcon2}
\eea
A consequence of the two equations above is that the right-hand-side can be rewritten in terms of the left-hand-side yielding (\ref{ccy}). As a result  the form $\tau^k$ satisfies the CKY equation.

Although CKY equations are suitable to describe the geometry of some manifolds that admit Killing spinors, as established above, the conditions imposed by generic KSEs, like those of supergravity theories, on the Killing spinor bi-linears   are far  more involved. Therefore a  suitable
generalization of the CKY condition is required to proceed further.

\subsection{Twisted form hierarchies and CKY equations}

  To give the definition of a
 TCFH, let $M$ be a manifold with a metric $g$ and  signature $(r,s)$, $\Lambda_c^*(M)$ be the complexified bundle of all forms on $M$ and ${\cal F}$ be a multi-form, i.e.  ${\cal F}$ be a collection of (complex) forms of non-necessarily different degree. In particular  ${\cal F}$ is a section of $\oplus^m\Lambda_c^*(M)$,  ${\cal F}\in \Gamma(\oplus^m\Lambda_c^*(M))$.  A  {\it TCFH } with ${\cal F}$  is
a collection of forms $\{\chi^p\}$  \cite{eigen}, with possibly different degrees $p$, which satisfy
\bea
\nabla^{\cal F}_X (\{\chi^p\})= i_X{\cal P}({\cal F}, \{\chi^p\})+ \alpha_X\wedge {\cal Q}({\cal F}, \{\chi^p\})~,
\label{covhie}
\eea
where ${\cal P}, {\cal Q}: \Gamma(\Lambda_c^*(M))\rightarrow \Gamma(\Lambda_c^*(M))$ and $\nabla^{\cal F}$, {\it the covariant hierarchy connection},  is a connection acting on  $\Gamma(\oplus^\ell\Lambda_c^*(M))$ constructed from the Levi-Civita
connection and ${\cal F}$.   In the application that follows, the multi-forms ${\cal P}$ and ${\cal Q}$  are  constructed  from $\{\chi^p\}$ and ${\cal F}$ via the use of algebraic operations like the  wedge product with ${\cal F}$, the inner derivation
with respect to ${\cal F}$ viewed as a multi-vector valued multi-form with indices raised with respect to the metric $g$, and their adjoints with respect to the standard inner product in $\Gamma(\Lambda_c^*(M))$.  The covariant hierarchy connection $\nabla^{\cal F}$ is not necessarily degree preserving.

In the applications to supergravity explored below  in sections \ref{msec} and \ref{esec}, ${\cal F}$ are the form field strengths of the theories and $\{\chi^p\}$ are the form bilinears constructed from a Killing spinor. In addition, the condition (\ref{covhie}) is analogous to the conditions (\ref{hcon1}) and (\ref{hcon2}) given in the previous section. Additional examples will be presented below in sections \ref{sec1}, \ref{sec2} and \ref{sec3}. There one can find     (\ref{covhie}) explicitly for  4-, 5- and 11-dimensional supergravities and so illustrate the definition\footnote{The definition of (\ref{covhie}) is rather involved because it has been designed to incorporate all TCFHs that can occur in supergravity theories. However for the examples presented, (\ref{covhie}) is given in a coordinate basis and so it is explicit.}.  The TCFHs have been used in \cite{eigen} to explore the geometry of manifolds that arise in certain eigenvalue estimates for a class of modified Dirac operators.  There is a further generalization of the TCFH by allowing a further twisting of $\oplus^\ell\Lambda_c^*(M)$ with a vector bundle $E$ and so $\nabla^{\cal F}$ becomes a connection acting on $\Gamma(\oplus^\ell\Lambda_c^*(M)\otimes E)$.  This generalization is required for gauged supergravities but we shall not elaborate on this. Though our main result is valid for the KSEs of these supergravities.

As we have seen (\ref{hcon1}) and (\ref{hcon2}) are associated with a CKY equation.  Indeed after comparing the left-hand and right-hand sides of (\ref{covhie}), one finds that
 \bea
 \big(\nabla_X^{\cal F}\{\chi_q\}\big)\vert_p={1\over p+1} \big(i_X d^{\cal F}(\{\chi_q\})\big)\vert_p-{1\over n-p+1} \alpha_X\wedge \big(\delta^{\cal F}(\{\chi_q\})\big)\vert_{p-1}~,
 \label{twky}
 \eea
 where $(\dots)\vert_p$ denotes a restriction of the expression to p-forms,  $d^{\cal F}$ is the exterior derivative constructed using $\nabla_X^{\cal F}$ and similarly $\delta^{\cal F}$ is an adjoint
 constructed using again $\nabla_X^{\cal F}$.  This follows immediately from   (\ref{covhie})  by shew-symmetrizing
 all indices first to derive the $d^{\cal F}$ term and then contracting with the metric to derive the $\delta^{\cal F}$ term. Incidentally this
  gives the definitions of  $d^{\cal F}$ and $\delta^{\cal F}$ operations.

 The equation (\ref{twky}) can be seen as a generalization of the CKY equations. Though there are several differences between (\ref{twky}) and previous generalizations of CKY equation. The new connection used in the relation
 $\nabla_X^{\cal F}$ may not degree preserving. Moreover (\ref{twky}) is a relation between a collection of forms with possibly different degrees while
 typically the standard CKY equation and other generalizations are conditions on a form with a definite degree.

 Note  that although (\ref{twky}) is implied by
  (\ref{covhie}), the converse it not necessarily the case unless one imposes in addition  that
 \bea
 {1\over p+1} \big(i_X d^{\cal F}(\{\chi_q\})\big)\vert_p=\big(i_X{\cal P}\big)\vert_p~,~~~-{1\over n-p+1}\big(\delta^{\cal F}(\{\chi_q\})\big)\vert_{p-1}={\cal Q}\vert_{p-1}~.
 \eea
Thus (\ref{twky}) is a coarser relation than that of (\ref{covhie}). Therefore, it is expected that there are solution of (\ref{twky})
which are not solutions of (\ref{covhie}).  Although of course all solutions of (\ref{covhie}) are also solutions of (\ref{twky}).

\section{Proof of main result} \label{msec}

To prove that the KSEs of a supergravity theory are associated with a TCFH, let us first recall that the supercovariant connection of a supergravity theory has the structure
\bea
{\cal D}_X=\nabla_X+c(i_X {\cal H})+ c(\alpha_X\wedge {\cal G})~,
\label{sup}
\eea
where $c$ denotes the Clifford algebra element associated with the multi-forms $i_X{\cal H}=\sum_p i_XH^p$, ${\cal G}=\sum_p G^p$, where $H^p$'s and $G^p$'s  are the
p-form field strengths of a supergravity theory. $\nabla_X$ is typically the Levi-Civita connection but it can also be twisted with a gauge connection.

Notice that we have not imposed any restrictions on the signature of the spacetime. The argument below applies to the KSEs of all standard supergravities defined on Lorentzian signature manifolds as well as to those of non-standard supergravities defined on  (r,s)-signature manifolds, e.g Euclidean signature manifolds.

As the argument that follows  is linear in the  field strengths $H^p$ and $G^p$, it is sufficient to take ${\cal H}$ and ${\cal G}$ to be  single forms $H$ and $G$ of degree $\ell$, $H^\ell=H$ and $G^\ell=G$.  Next consider the form bi-linears $\{\chi^p\}$ constructed from a Killing spinors $\epsilon$, ${\cal D}_X\epsilon=0$, with respect to some Spin-invariant inner product say $\langle\cdot, \cdot\rangle_s$, i.e.
 \bea
 \chi^p={1\over p!} \langle\epsilon, \Gamma_{A_1\dots A_p}\epsilon\rangle_s\, e^{A_1}\wedge\dots \wedge e^{A_p}~,
 \label{bibic}
 \eea
 where $\{e^A\}$ is a (pseudo)-orthonormal frame adapted to the spacetime metric.
 See e.g. appendix B in \cite{rev} for a discussion on Spin-invariant bi-linears.  Then one has that
\bea
&&\nabla_X \chi^p=-{1\over p!}\big(\langle c(i_X  H)\epsilon, \Gamma_{A_1\dots A_p}\epsilon\rangle_s+\langle \epsilon, \Gamma_{A_1\dots A_p}c(i_X  H)\epsilon\rangle_s\big)\, e^{A_1}\wedge\dots \wedge e^{A_p}
\cr
&&
\quad-{1\over p!}\big(\langle c(\alpha_X\wedge G)\epsilon, \Gamma_{A_1\dots A_p}\epsilon\rangle_s+\langle \epsilon, \Gamma_{A_1\dots A_p}c(\alpha_X\wedge G)\epsilon\rangle_s\big)\, e^{A_1}\wedge\dots \wedge e^{A_p}~.
\label{covdev}
\eea
After using the Hermiticity properties of the inner product and the Clifford algebra relation as well as the definition of the form bi-linears, one finds that
\bea
&&{1\over p!}(\langle c(i_X  H)\epsilon, \Gamma_{A_1\dots A_p}\epsilon\rangle_s+\langle \epsilon, \Gamma_{A_1\dots A_p}c(i_X  H)\epsilon\rangle_s)\, e^{A_1}\wedge\dots \wedge e^{A_p}
\cr
&&\qquad=\bigg(\sum_q (c^1_q\, i_XH\cdot \chi^q+\tilde c^1_q\, i_X\bar H\cdot \chi^q)\bigg)\vert_p~,
\label{xxyy}
\eea
where $\bar H$ is the complex conjugate of $H$,  $c^1_q$ and $\tilde c^1_q$ are combinatorial coefficients which depend on $p$, $\ell$ and the inner product $\langle\cdot, \cdot\rangle_s$ whose values are not essential for the proof that follows. Although they are computed explicitly in the examples
presented below.  Moreover $i_XH\cdot \chi^q$ denotes multi-index contractions between the $i_XH$ and $\chi^q$ forms  and similarly for $i_X\bar H\cdot \chi^q$,
where the indices are raised with respect to the metric.  For example $\psi^k\cdot \omega^m$ denotes any of the contractions
\bea
(\psi^k\cdot \omega^m)_{N_1\dots N_p}={1\over s!} \psi^{M_1\dots M_s}{}_{[N_1\dots N_{k-s}} \omega_{|M_1\dots M_s|N_{k-s+1}\dots N_p]}~,
\eea
 of the forms $\psi^k$ and $\omega^m$ and $p=k+m-2s$.

Furthermore, one can show after using the Clifford algebra relations and the properties of $\langle\cdot, \cdot\rangle_s$ that
\bea
&&{1\over p!}(\langle c(\alpha_X\wedge G)\epsilon, \Gamma_{A_1\dots A_p}\epsilon\rangle_s+\langle \epsilon, \Gamma_{A_1\dots A_p}c(\alpha_X\wedge G)\epsilon\rangle_s)\, e^{A_1}\wedge\dots \wedge e^{A_p}
\cr
&&\qquad =\alpha_X\wedge\bigg(\sum_q  (c^2_q\, G\cdot \chi^q+\tilde c^2_q\, \bar G\cdot \chi^q)\bigg)\vert_{p-1}
\cr
&&\qquad\qquad+\bigg(\sum_q
(c^3_q\, G\cdot i_X\chi^q+\tilde c^3_q\, \bar G\cdot i_X\chi^q)\bigg)\vert_p~.
\eea
 where again the $c$'s are combinatorial coefficients that their value is not essential for the proof. Note that in the above expression
the terms  $i_XG\cdot \chi^q$ and $i_X\bar G\cdot \chi^q$ do not contribute. This is one of the key observations required for the proof of the statement and it is a consequence of the Clifford algebra relation. The last term in the above expression can be rewritten as
 \bea
 \bigg(\sum_q
(c^3_q\, G\cdot i_X\chi^q+\tilde c^3_q\, \bar G\cdot i_X\chi^q)\bigg)\vert_p&=&\bigg(i_X \big(\sum_q (c^4_q\, G\cdot \chi^q+\tilde c^4_q\, \bar G\cdot \chi^q)\big)\bigg)\vert_p
\cr
&&\quad +\bigg(\sum_q (c^5_q\, i_XG\cdot \chi^q+\tilde c^5_q\, i_X\bar G\cdot \chi^q)\bigg)\vert_p~,
\eea
for some combinatorial coefficients $c^4_q, c^5_q$ and $\tilde c^4_q, \tilde c^5_q$.
 The last term in the above expression has the same structure as that in (\ref{xxyy}). Using this,  (\ref{covdev}) can be written after some rearrangement as
 \bea
 \nabla_X \chi_p&+&\bigg(\sum_q \big((c^1_q\, i_XH+c^5_q i_XG)\cdot \chi^q+(\tilde c^1_q\, i_X\bar H+\tilde c^5_q\, i_XG)\cdot \chi^q\big )\bigg)\vert_p=
 \cr
 && \qquad-\bigg(i_X \big(\sum_q (c^4_q\, G\cdot \chi^q+\tilde c^4_q\, \bar G\cdot \chi^q)\big)\bigg)\vert_p
 \cr
 &&\qquad -\alpha_X\wedge\bigg(\sum_q  (c^2_q\, G\cdot \chi^q+\tilde c^2_q\, \bar G\cdot \chi^q)\bigg)\vert_{p-1}~.
 \label{covhie2}
 \eea
 Clearly this equation defines a TCFH (\ref{covhie})
 with
 \bea
&& \nabla_X^{\cal F}\chi_p\equiv \nabla_X \chi_p+\bigg(\sum_q \big((c^1_q\, i_XH+c^5_q i_XG)\cdot \chi^q+(\tilde c^1_q\, i_X\bar H+\tilde c^5_q\, i_XG)\cdot \chi^q\big )\bigg)\vert_p~,
 \cr
 &&\big(i_X{\cal P}\big)\vert_p\equiv -\bigg(i_X \big(\sum_q (c^4_q\, G\cdot \chi^q+\tilde c^4_q\, \bar G\cdot \chi^q)\big)\bigg)\vert_p~,
 \cr
 &&
 {\cal Q}\vert_{p-1}\equiv-\bigg(\sum_q  (c^2_q\, G\cdot \chi^q+\tilde c^2_q\, \bar G\cdot \chi^q)\bigg)\vert_{p-1}~,
 \eea
 and ${\cal F}=\{{\cal H}, {\cal G}\}_{\mathrm{ind}}$ are the linearly independent form field strengths.
 Note that $\nabla_X^{\cal F}$ as defined above is a connection in the space of forms-it satisfies all four  axioms of a connection-but it is not necessarily degree preserving unless all combinatorial coefficients $c^0_q$ and $\tilde c^0_q$ vanish apart from those with $q=p$.  In particular observe that it satisfies the linearity properties $\nabla_{X+Y}^{\cal F}=\nabla_{X}^{\cal F}+\nabla_{Y}^{\cal F}$ and $\nabla_{X}^{\cal F}(\{\chi^p+\psi^p\})=\nabla_{X}^{\cal F}(\{\chi^p\}) +\nabla_{X}^{\cal F}(\{\psi^p\})$ as well as $\nabla_{fX}^{\cal F}= f \nabla_{X}^{\cal F}$ and the Leibniz type of rule $\nabla_{X}^{\cal F}(\{f\chi^p\})= X(f) \{\chi^p\}+ f\nabla_{X}^{\cal F}(\{\chi^p\})$, where $f$ is a function on $M$.  This completes the proof of the main result.

 The proof above utilizes  the form bi-linears  (\ref{bibic}) of a single Killing spinor $\epsilon$. However, it can be easily generalized to include the bi-linears of any number of Killing spinors with respect
 to any Spin-invariant inner product. The proof is essentially the same. The only difference is that instead of considering the forms in (\ref{bibic}), one should replace them with
  \bea
 \chi^p={1\over p!} \langle\eta, \Gamma_{A_1\dots A_p}\epsilon\rangle_s\, e^{A_1}\wedge\dots \wedge e^{A_p}~,
 \label{bibicx}
 \eea
where both $\eta$ and $\epsilon$ are Killing spinors.

The theorem can also be generalized to effective  theories that include supergravity with higher order corrections, like for example those that emerge
as low energy effective theories of superstrings and M-theory. This is because the general structure of the supercovariant connection after including higher order corrections
is expected to have the general form of (\ref{sup}).  In fact for the theorem to apply, it is not necessary to assume that ${\cal H}$ is a form. Instead ${\cal H}$ can be a section of $\Lambda^1_c(M)\times \Lambda^*_c(M)$ provided    that $i_X{\cal H}$ which appears in the expression for the  supercovariant derivative (\ref{sup}) is a multi-form.

 It should also be  noted that there is an ambiguity in the definition of $\nabla^{\cal F}$.  To see this observe that there may be terms
 ${\cal F}\cdot \chi_p$ which have the property that
 \bea
 i_X({\cal F}\cdot \chi_p)=i_X{\cal F}\cdot \chi_p~,
 \eea
 e.g. terms for which all indices of $\chi_p$ are contracted to indices of ${\cal F}$.  Such terms can either  contribute to $\nabla^{\cal F}$ or to ${\cal P}$.  If all such terms are included in  $\nabla^{\cal F}$, then  such a covariant hierarchy connection will be called {\it maximal} and denoted with $\nabla^{\cal F}$. On the other hand   if all such terms are included in ${\cal P}$, then   $\nabla^{\cal F}$ will be called  {\it minimal} and denoted with ${\cal D}^{\cal F}$. It is clear that there may be many intermediate cases.

 The TCFHs that arise in supergravity theory can always be chosen such that the Hodge duality operation on $\{\chi^p\}$ is an automorphism of the hierarchy.
 This can always be achieved by choosing in the set $\{\chi^p\}$ all bi-linears and their Hodge duals. In such a case the hierarchy will be twisted by ${\cal F}$ as
 originally has been indicated. However in many of the examples below the set of bi-linears $\{\chi^p\}$  is chosen up to a Hodge duality operation. This simplifies
 the selection and so the final result.  In such a case, the Hodge duality operation may not be an automorphism of the TCFH.  In addition such a  TCFH will be twisted
 with respect to both ${\cal F}$ and its dual ${}^*{\cal F}$.

Furthermore, if the fluxes are chosen such that the supercovariant connection ${\cal D}_X$ depends only on $i_X{\cal F}$, then there is a choice of $\{\chi^p\}$ such that the associated TCFH is a parallel transport equation with respect to $\nabla^{\cal F}$ connection\footnote{It would be of interest to explore the relation of such a $\nabla^{\cal F}$ connection to the supercovariant connection ${\cal D}_X$ acting on the tensor product ${\cal S}\otimes{\cal S}$ of two spinor bundles ${\cal S}$ which can be converted to an  action on the form  spinor bilinears via Fierz identities .}. In such and case ${\cal P}={\cal Q}=0$. To achieve this, the basis chosen for the fluxes and form bilinears will appropriately include the fluxes and/or their Hodge duals.    In any case the theorem proven above demonstrates that whatever the choice of basis in the fluxes and form spinor bilinears is, the KSEs of supergravity theories give rise to a TCFH and in turn to a generalization of CKY equations.

\section{Examples}\label{esec}

To illustrate the proof given above, we shall present some examples and explore their properties. It is not the  purpose here to give a complete description of the TCFHs of  all supergravity theories-this
 will be presented elsewhere and it will include the 10-dimensional type II supergravity theories that arise in the context of string theory. To begin the supercovariant connection of the heterotic   and ungauged $N=(1,0)$ $d=6$  supergravity theories is induced from the connection
 on the tangent bundle of the spacetime which has skew-symmetric torsion $H$.  Denoting this connection with $\nabla^H=\nabla+{1\over2}H$, it is straightforward to observe  that the twisted covariant hierarchy  connection $\nabla^{\cal F}$ coincides with $\nabla^H$.  Furthermore the Killing spinor bi-linears are $\nabla^H$-covariantly constant.

 The same conclusion holds for the heterotic strings for up and including 2-loop corrections in the sigma model perturbation theory.
In the gauged $N=(1,0)$ $d=6$ supergravity some of the form Killing spinor bi-linears are twisted with respect to the gauge connection. Again $\nabla^{\cal F}$ is given by $\nabla^{\cal F}=\nabla^H+ A^r D_r$, where $A^r$ is the gauge connection and $D_r$ is the representation
 of the  Lie algebra of the gauge group acting on some of the form-bi-linears. The holonomy of these connections has been explored in \cite{het1, het2}  to classify all the supersymmetric solutions of heterotic theory  and those of $N=(1,0)$ $d=6$ \cite{gutowski, akyol} supergravities coupled to matter multiplets.

\subsection{Minimal $N=2$ supergravity in four dimensions} \label{sec1}
 As another example consider the $N=2$ $d=4$ minimal supergravity. To begin  the bi-linears constructed from the
 Dirac inner product, $\langle\cdot,\cdot\rangle_D$,  are
 \bea
 &&f=\langle \epsilon, \epsilon\rangle_D~,~~~K=\langle \epsilon, \Gamma_A\epsilon\rangle_D\, e^A~,~~~\omega={1\over2}\langle \epsilon, \Gamma_{AB}\epsilon\rangle_D\, e^A\wedge e^B~,~~~
 \cr
 &&Y=\langle \epsilon, \Gamma_A \gamma_5\epsilon\rangle_D\, e^A~,~~~g=\langle \epsilon, \gamma_5\epsilon\rangle_D~,
 \label{bibi}
 \eea
 where $\gamma_5=i\Gamma_{0123}$.  These are the algebraically independent bi-linears up to a Hodge duality operation.
 The supercovariant derivative of the theory is
 \bea
 {\cal D}_M\equiv \nabla_M+{i\over4} F_{AB} \Gamma^{AB} \Gamma_M~.
 \eea
 Assuming the $\epsilon$ is a Killing spinor, ${\cal D}_M\epsilon=0$, one can easily compute the covariant derivative of the bi-linear and re-arrange the terms
 as a TCFH.  In particular, the TCFH with respect to the minimal connection reads
 \bea
&& {\cal D}_M^{\cal F}f\equiv \nabla_M f=i K^A F_{MA}~,~~~{\cal D}_M^{\cal F}K_N\equiv \nabla_M K_N=i f F_{MN}-g\, {}^*F_{MN}~,~~~
\cr
&&{\cal D}_M^{\cal F}\omega_{NR}\equiv \nabla_M\omega_{NR}-4\,{}^*F_{M[N} Y_{R]}= -3\, {}^*F_{[MN} Y_{R]}-2 g_{M[N} {}^*F_{R]D} Y^D~,~~~
\cr
&&
{\cal D}_M^{\cal F} Y_N\equiv \nabla_M Y_N+{}^*F_{MA} \omega^A{}_N=-{1\over2} g_{MN}\, {}^*F_{PQ} \omega^{PQ}+{}^*F_{[M|A|}\omega^A{}_{N]}
~,~~~
\cr
&&
{\cal D}_M^{\cal F} g\equiv \nabla_Mg={}^*F_{MN}K^N~,
 \eea
 where ${}^*F_{MN}={1\over2}\epsilon_{MNPQ} F^{PQ}$ with $\epsilon_{0123}=-1$.  Observe that the above equations can be arranged to be real with an appropriate
 redefinition of the form bi-linears as in the basis chosen some of them are imaginary.  Note also that in the computation for the 2-form $\omega$ a term arises
 with the structure $(\alpha_X\wedge {}^*F)\cdot Y$ and it has been rewritten as a linear combination of $Y\wedge i_X{}^*F$, $\alpha_X\wedge {}^*F\cdot Y$ and $i_X(Y\wedge {}^*F)$.

 Similarly the TCFH with respect to
 maximal connection is
 \bea
&& \nabla^{\cal F}_M f\equiv \nabla_M f-i K^A F_{MA}=0~,~~~\nabla^{\cal F}_M  K_N\equiv \nabla_M K_N-i f F_{MN}+g\, {}^*F_{MN}=0~,~~~
\cr
&&
\nabla^{\cal F}_M \omega_{NR}\equiv \nabla_M\omega_{NR}-4\,{}^*F_{M[N} Y_{R]}= -3\, {}^*F_{[MN} Y_{R]}-2 g_{M[N} {}^*F_{R]D} Y^D~,
 \cr
 &&
 \nabla^{\cal F}_M Y_N\equiv \nabla_M Y_N+{}^*F_{MA} \omega^A{}_N=-{1\over2} g_{MN}\, {}^*F_{PQ} \omega^{PQ}+{}^*F_{[M|A|}\omega^A{}_{N]}~,~~~
 \cr &&
 \nabla^{\cal F}_M g\equiv \nabla_Mg-{}^*F_{MN}K^N=0~.
 \eea
It is clear that in both cases the covariant form hierarchy is twisted by the multi-form ${\cal F}=\{F, {}^*F\}$.  Note that if the basis in the
space of bi-linears included the Hodge dual forms, then the hierarchy would have been twisted just with ${\cal F}=\{F \}$.

The associated generalized   CKY equations with respect to the minimal connection are
\bea
&& {\cal D}^{\cal F}_Mf={\cal D}^{\cal F}_Mf~,~~~{\cal D}^{\cal F}_M K_N={\cal D}^{\cal F}_{[M}K_{N]}~,~~~
{\cal D}^{\cal F}_M\omega_{NR}= {\cal D}^{\cal F}_{[M}\omega_{NR]}+{2\over3}g_{M[N}{\cal D}^{\cal F}{}^P\omega_{|P|R]}~,~~~
\cr
&&
{\cal D}^{\cal F}_M Y_N={1\over4} g_{MN}\, g^{PQ} {\cal D}^{\cal F}_P Y_Q +{\cal D}^{\cal F}_{[M} Y_{N]}
~,~~~
{\cal D}^{\cal F}_M g={\cal D}^{\cal F}_M g~.
 \eea
 The associated generalized   CKY equations with respect to the maximal  connection are
 \bea
&&\nabla^{\cal F}_Mf=0~,~~~\nabla^{\cal F}_M K_N=0~,~~~\nabla^{\cal F}_M\omega_{NR}= \nabla^{\cal F}_{[M}\omega_{NR]}+{2\over3}g_{M[N}{\nabla}^{\cal F}{}^P\omega_{|P|R]}~,
 \cr
&&
\nabla^{\cal F}_M Y_N={1\over4} g_{MN}\, g^{PQ} \nabla^{\cal F}_P Y_Q +\nabla^{\cal F}_{[M} Y_{N]}~,~~~
\nabla^{\cal F}_M g=0~.
 \eea
Observe that if for some background ${}^*F_{M[N} Y_{R]}=0$, then $\omega$ satisfies the KY equation. Of course the vector field associated to $K$ is Killing as expected.

 \subsection{Minimal $N=1$ supergravity in five dimensions} \label{sec2}

 Next let us turn to $N=1$  supergravity in five dimensions. Consider the algebraically independent Killing spinor bi-linears up to a Hodge duality operation
  \bea
 &&f=\langle \epsilon, \epsilon\rangle_D~,~~~K=\langle \epsilon, \Gamma_A\epsilon\rangle_D\, e^A~,~~~\omega={1\over2}\langle \epsilon, \Gamma_{AB}\epsilon\rangle_D\, e^A\wedge e^B~,~~~
 \label{biib}
 \eea
 where now $\epsilon$ is a $\mathrm{Spin}(4,1)$ spinor and $\Gamma_4=\Gamma^4=i\Gamma_{0123}$. The supercovariant connection of the theory is
 \bea
 {\cal D}_M\equiv \nabla_M-{i\over 4\sqrt3}\big(\Gamma_M{}^{AB}F_{AB}-4 F_{MA} \Gamma^A\big)~.
 \eea
 Assuming the $\epsilon$ is a Killing spinor, ${\cal D}_M\epsilon=0$,  the conditions on the bi-linears imposed by the gravitino KSE have been found in \cite{pakis}.
 These conditions have been given in a non-TCFH expression.  Putting them into   the TCFH form with respect to the minimal connection, one finds  that
 \bea
 &&{\cal D}_M^{\cal F}f\equiv \nabla_Mf=-{2i\over\sqrt3} F_{MN} K^N~,~~~
 \cr
 &&
 {\cal D}_M^{\cal F} K_N\equiv \nabla_M K_N={1\over \sqrt3}{}^*F_{MNR} K^R-{2i\over\sqrt3} F_{MN}f~,~~~
 \cr
 &&
 {\cal D}_M^{\cal F}\omega_{NR}\equiv \nabla_M\omega_{NR}-\sqrt3\, {}^*F_{MNE} \omega^E{}_R+\sqrt3\, {}^*F_{MRE} \omega^E{}_N
 \cr
 \qquad &&=-2\sqrt3\, {}^*F_{E[NR} \omega^E{}_{M]}+ {2\over\sqrt3} g_{M[N}\, {}^*F_{R]EF} \omega^{EF}~.
 \eea
Similarly, the TCFH with respect to the maximal connection is
\bea
 &&\nabla^{\cal F}_Mf\equiv \nabla_Mf+{2i\over\sqrt3} F_{MN} K^N=0~,~~~
 \cr
 &&
 \nabla^{\cal F}_M K_N\equiv \nabla_M K_N-{1\over \sqrt3}{}^*F_{MNR} K^R+{2i\over\sqrt3} F_{MN}f=0~,~~~
 \cr
 &&
 \nabla^{\cal F}_M\omega_{NR}\equiv \nabla_M\omega_{NR}-\sqrt3\, {}^*F_{MNE}\, \omega^E{}_R+\sqrt3\, {}^*F_{MRE}\, \omega^E{}_N
 \cr
 \qquad &&=-2\sqrt3\, {}^*F_{E[NR} \omega^E{}_{M]}+ {2\over\sqrt3} g_{M[N}\, {}^*F_{R]EF}\, \omega^{EF}~.
 \eea
 Note that there are additional form bi-linears that can be added to (\ref{biib}) as the theory always preserves even number of supersymmetries. However
 the choice made above suffices to demonstrate the general theorem.

The associated generalizations of the CKY equations are
\bea
&&{\cal D}^{\cal F}_Mf={\cal D}^{\cal F}_Mf~,~~~{\cal D}^{\cal F}_M K_N={\cal D}^{\cal F}_{[M} K_{N]}~,
\cr
&&
{\cal D}^{\cal F}_M\omega_{NR}  ={\cal D}^{\cal F}_{[M} \omega_{NR]}
-{1\over2}g_{M[N} {\cal D}^{\cal F}{}^E\omega_{R]E}~,
\eea
and
\bea
 \nabla^{\cal F}_Mf=0~,~~~\nabla^{\cal F}_M K_N=0~,~~~
\nabla^{\cal F}_M\omega_{NR}=\nabla^{\cal F}_{[M} \omega_{NR]}
-{1\over2}g_{M[N} \nabla^{\cal F}{}^E\omega_{R]E}~,
\eea
respectively.  Observe that both the minimal and maximal TCFH connections on $\omega$ are connections with skew-symmetric torsion $H={2\over\sqrt 3} {}^*F$.  In turn the associated
generalization of the CKY equation is that for which the Levi-Civita connection $\nabla$ is replaced with $\nabla^H$.  As a result {\it all supersymmetric solutions} of $N=1$ $d=5$ supergravity
with $\omega\not=0$ admit a CKY 2-form associated with a connection with skew-symmetric torsion.  This includes electrically and magnetically  charged black holes as well pp-wave backgrounds. Furthermore again $K$ is associated with a Killing vector field.

\subsection{11-dimensional supergravity} \label{sec3}
As a final example, let us consider the $N=1$ supergravity in eleven dimensions and backgrounds that admit one Killing spinor. Such backgrounds admit a 1-form, $K$, 2-form, $\omega$, and 5-form, $\tau$, Killing spinor bi-linears up to a Hodge duality operation.
The supercovariant connection of 11-dimensional supergravity is
 \bea
 {\cal D}_M\equiv \nabla_M+{1\over288}\big( \Gamma_M{}^{N_1N_2N_3N_4} F_{N_1N_2N_3N_4}-8 F_{MN_1N_2N_3} \Gamma^{N_1N_2N_3}\big)~.
 \eea
 Using ${\cal D}_M\epsilon=0$, the covariant derivative of these bi-linears has been computed in \cite{pakis2}. Again the equations are not in TCFH form.  Rewriting the expressions as a TCFH with respect to the minimal connection these read as
\bea
&&{\cal D}^{\cal F}_M K_N\equiv \nabla_M K_N={1\over6} F_{MNPQ}\, \omega^{PQ}-{1\over 6!}\, {}^*F_{MNP_1\dots P_5}\, \tau^{P_1\dots P_5}~,
\cr
&&
{\cal D}^{\cal F}_M\omega_{NR}\equiv \nabla_M\omega_{NR}-{1\over 2\cdot 3!} F_{ME_1E_2E_3}\, \tau^{E_1E_2E_3}{}_{NR}=-{1\over 3} F_{MNRE} K^E
\cr
&&
\qquad
- {1\over 2\cdot 3!} \tau_{[MN}{}^{E_1E_2E_3} F_{R]E_1E_2E_3}+{1\over 3\cdot 4!} g_{M[N}\,\tau_{R]}{}^{E_1\dots E_4} F_{E_1\dots E_4}~,
\cr
&&
{\cal D}^{\cal F}_M \tau_{N_1\dots N_5}\equiv \nabla_M\tau_{N_1\dots N_5}+5 F_{M[N_1N_2N_3}\, \omega_{N_4N_5]}-{5\over6} {}^*F_{M[N_1N_2N_3|E_1E_2E_3|}\, \tau_{N_4N_5]}{}^{E_1E_2E_3}=
\cr
&&\qquad -{1\over6}{}^*F_{MN_1\dots N_5E}K^E+{5\over2} F_{[MN_1N_2N_3}\, \omega_{N_4N_5]}-{5\over6} \tau_{[MN_1}{}^{E_1E_2E_3}\, {}^*F_{N_3\dots N_5]E_1E_2E_3}
\cr
&&
-{10\over3} g_{M[N_1} \omega^E{}_{N_2} F_{N_3N_4N_5]E}
 -{5\over18} g_{M[N_1}\, \tau_{N_2}{}^{E_1E_2E_3E_4}\, {}^*F_{N_3N_4N_5] E_1E_2E_3E_4}~,
\eea
where $\epsilon_{01\dots9(10)}=-1$\footnote{Our convention for the Levi-Civita tensor differs from that in \cite{pakis2}.}.
While the TCFH  with respect to the maximal connection is
\bea
&&\nabla^{\cal F}_M K_N\equiv\nabla_M K_N-{1\over6} F_{MNPQ}\, \omega^{PQ}+{1\over 6!}\, {}^*F_{MNP_1\dots P_5}\, \tau^{P_1\dots P_5}=0~,
\cr
&&
\nabla^{\cal F}_M\omega_{NR}\equiv\nabla_M\omega_{NR}-{1\over 2\cdot 3!} F_{ME_1E_2E_3}\, \tau^{E_1E_2E_3}{}_{NR}+{1\over 3} F_{MNRE} K^E=
\cr
&&
\qquad
- {1\over 2\cdot 3!} \tau_{[MN}{}^{E_1E_2E_3} F_{R]E_1E_2E_3}+{1\over 3\cdot 4!} g_{M[N}\,\tau_{R]}{}^{E_1\dots E_4} F_{E_1\dots E_4}~,
\cr
&&
\nabla^{\cal F}_M \tau_{N_1\dots N_5}\equiv \nabla_M\tau_{N_1\dots N_5}+5 F_{M[N_1N_2N_3}\, \omega_{N_4N_5]}-{5\over6} {}^*F_{M[N_1N_2N_3|E_1E_2E_3|}\, \tau_{N_4N_5]}{}^{E_1E_2E_3}
\cr
&&\qquad +{1\over6}{}^*F_{MN_1\dots N_5E}K^E={5\over2} F_{[MN_1N_2N_3}\, \omega_{N_4N_5]}-{5\over6} \tau_{[MN_1}{}^{E_1E_2E_3}\, {}^*F_{N_3\dots N_5]E_1E_2E_3}
\cr
&&
-{10\over3} g_{M[N_1}\, \omega^E{}_{N_2} F_{N_3N_4N_5]E}
 -{5\over18} g_{M[N_1}\, \tau_{N_2}{}^{E_1E_2E_3E_4}\, {}^*F_{N_3N_4N_5] E_1E_2E_3E_4}~.
\eea
Clearly the TCFHs are twisted with respect to ${\cal F}=\{F, {}^*F\}$.

The associated twisted CKY equations are
\bea
&&{\cal D}^{\cal F}_M K_N={\cal D}^{\cal F}_{[M} K_{N]}~,~~~
{\cal D}^{\cal F}_M\omega_{NR}= {\cal D}^{\cal F}_{[M}\omega_{NR]} -{1\over5}g_{M[N} {\cal D}^{\cal F}{}^E\omega_{R]E}~,
\cr
&&
{\cal D}^{\cal F}_M \tau_{N_1\dots N_5}= {\cal D}^{\cal F}_{[M} \tau_{N_1\dots N_5]}+{5\over7} g_{M[N_1} {\cal D}^{\cal F}{}^E\,\tau_{N_2\dots N_5]E}~,
\eea
and
\bea
&&\nabla^{\cal F}_M K_N=0~,~~~
\nabla^{\cal F}_M\omega_{NR}= \nabla^{\cal F}_{[M}\omega_{NR]} -{1\over5}g_{M[N} \nabla^{\cal F}{}^E\omega_{R]E}~,
\cr
&&
\nabla^{\cal F}_M \tau_{N_1\dots N_5}= \nabla^{\cal F}_{[M} \tau_{N_1\dots N_5]}+{5\over7} g_{M[N_1} \nabla^{\cal F}{}^E\,\tau_{N_2\dots N_5]E}~,
\eea
respectively.   $K$ is associated with a Killing vector field.  It is expected that in many special backgrounds the bi-linears will satisfy the CKY equations as some components of the fluxes will vanish and so the TCFH connections will be simplified.

\section*{Conclusions}

 We have demonstrated that the gravitino KSE of all supergravity theories, irrespective of spacetime signature and including higher order corrections, are associated with a family of TCFHs. Each TCFH
 in the family is characterized with a connection $\nabla_X^{\cal F}$ on the space of forms which may not be degree preserving. In turn each TCFH  gives rise to a twisted CKY structure on all supersymmetric solutions of a supergravity theory.
 This result establishes  a close relation between KSEs and  suitable generalizations of the CKY equation.

 It is clear that the  TCFH connections $\nabla_X^{\cal F}$ characterize the underlying geometric structure of a supersymmetric background of a supergravity theory.
 It is likely that their holonomy group for a generic background  is of general linear type  in analogy with the holonomy of the
 supercovariant connections \cite{hull, tsimpis, duff}.  As there is a $\nabla_X^{\cal F}$ connection for  each member in the family of  TCFHs associated with a supergravity theory, one expects that the holonomy of the connections $\nabla_X^{\cal F}$ is a refinement of that of the supercovariant connection. Some understanding  in this direction may be obtained by computing the holonomy of the $\nabla_X^{\cal F}$ for some well-known supersymmetric backgrounds.  In turn this may give some insight into the symmetries of M-theory.  It may also turn out that supersymmetric backgrounds can be characterized
 with the holonomy of TCFH connections.

 The proof presented here that all supersymmetric solutions admit a suitable generalization of the CKY equation opens another avenue towards understanding
 the geometric structure of supersymmetric backgrounds. It is likely that for many special supersymmetric backgrounds the twisted CKY equations simplify
 to the more standard CKY equations possibly twisted with a connection which is form degree preserving. Furthermore the geometry of supersymmetric backgrounds
 can be understood in parallel with that of some  non-supersymmetric ones that have been known for sometime that they admit CKY forms, for a review see \cite{frolov}.

 In a similar theme, as the twisted CKY condition is more coarse than that of an associated TCFH, it is likely that there are solutions of the twisted CKY
 equation that are not solutions of the  TCFH condition.  Such non-supersymmetric solutions will be very closely related to supersymmetric ones. It would be of interest
 to construct examples of such backgrounds.

 It is well known the  CKY equation is associated with the conserved charges of supersymmetric relativistic and non-relativist particle actions. Therefore the question arises whether this is the case for the twisted CKY equations associated with supersymmetric backgrounds found in this work.  The construction
 of such particle or possibly string actions that exhibit such symmetries may give some insights into these theories.

\section*{Acknowledgments}

I would like to thank Jan  Gutowski for many helpful discussions.


\begin{thebibliography}{99}


\bibitem{rev}
  U.~Gran, J.~Gutowski and G.~Papadopoulos,
  ``Classification, geometry and applications of supersymmetric backgrounds,''
  Phys.\ Rept.\  {\bf 794} (2019) 1
  doi:10.1016/j.physrep.2018.11.005
  [arXiv:1808.07879 [hep-th]].


\bibitem{pakis}
  J.~P.~Gauntlett, J.~B.~Gutowski, C.~M.~Hull, S.~Pakis and H.~S.~Reall,
  ``All supersymmetric solutions of minimal supergravity in five- dimensions,''
  Class.\ Quant.\ Grav.\  {\bf 20} (2003) 4587
  doi:10.1088/0264-9381/20/21/005
  [hep-th/0209114].


\bibitem{spingeom}
  J.~Gillard, U.~Gran and G.~Papadopoulos,
  ``The Spinorial geometry of supersymmetric backgrounds,''
  Class.\ Quant.\ Grav.\  {\bf 22} (2005) 1033
  doi:10.1088/0264-9381/22/6/009
  [hep-th/0410155].


  \bibitem{eigen}
  J.~Gutowski and G.~Papadopoulos,
  ``Eigenvalue estimates for multi-form modified Dirac operators,''
  arXiv:1911.02281 [math.DG].



  \bibitem{penrose}
R. Penrose, Ann. N.Y. Acad. Sci. 224 (1973) 125.


\bibitem{floyd}
  R. Floyd, The dynamics of Kerr fields,. Ph. D. Thesis, London (1973).

  \bibitem{carter-a}
  B.~Carter,
  ``Killing Tensor Quantum Numbers And Conserved Currents In Curved Space,''
  Phys.\ Rev.\  D {\bf 16} (1977) 3395.

  \bibitem{carter-b}
  B.~Carter,
  ``Global structure of the Kerr family of gravitational fields,''
  Phys.\ Rev.\  {\bf 174} (1968) 1559.

  \bibitem{chandrasekhar}
  S.~Chandrasekhar,
  ``The Solution Of Dirac's Equation In Kerr Geometry,''
  Proc.\ Roy.\ Soc.\ Lond.\  A {\bf 349} (1976) 571.

  \bibitem{carter-c}
  B.~Carter and R.~G.~Mclenaghan,
  ``Generalized Total Angular Momentum Operator For The Dirac Equation In
  Curved Space-Time,''
  Phys.\ Rev.\  D {\bf 19} (1979) 1093.

  \bibitem{page}
  P.~Krtous, D.~Kubiznak, D.~N.~Page and V.~P.~Frolov,
  ``Killing-Yano tensors, rank-2 Killing tensors, and conserved quantities in
  higher dimensions,''
  JHEP {\bf 0702} (2007) 004
  [arXiv:hep-th/0612029].

\bibitem{lun}
  Y.~Chervonyi and O.~Lunin,
  ``Killing(-Yano) Tensors in String Theory,''
  JHEP {\bf 1509} (2015) 182
  doi:10.1007/JHEP09(2015)182
  [arXiv:1505.06154 [hep-th]].





  \bibitem{revky}
  M.~Cariglia,
  ``Hidden Symmetries of Dynamics in Classical and Quantum Physics,''
  Rev.\ Mod.\ Phys.\  {\bf 86} (2014) 1283
  doi:10.1103/RevModPhys.86.1283
  [arXiv:1411.1262 [math-ph]].


\bibitem{frolov}
V.~Frolov, P.~Krtous and D.~Kubiznak,
``Black holes, hidden symmetries, and complete integrability,''
Living Rev. Rel. \textbf{20} (2017) no.1, 6
doi:10.1007/s41114-017-0009-9
[arXiv:1705.05482 [gr-qc]].




  \bibitem{gibbons}
  G.~W.~Gibbons, R.~H.~Rietdijk and J.~W.~van Holten,
  ``SUSY in the sky,''
  Nucl.\ Phys.\  B {\bf 404} (1993) 42
  [arXiv:hep-th/9303112].


  \bibitem{sfetsos}
  F.~De Jonghe, K.~Peeters and K.~Sfetsos,
  ``Killing-Yano supersymmetry in string theory,''
  Class.\ Quant.\ Grav.\  {\bf 14} (1997) 35
  [arXiv:hep-th/9607203].

  \bibitem{hl}
  P.~S.~Howe and U.~Lindström,
  ``Some remarks on (super)-conformal Killing-Yano tensors,''
  JHEP {\bf 1811} (2018) 049
  doi:10.1007/JHEP11(2018)049
  [arXiv:1808.00583 [hep-th]].

  \bibitem{warnick1}
  T.~Houri, D.~Kubiznak, C.~M.~Warnick and Y.~Yasui,
  ``Generalized hidden symmetries and the Kerr-Sen black hole,''
  JHEP {\bf 1007} (2010) 055
  [arXiv:1004.1032 [hep-th]].


\bibitem{warnick2}
  T.~Houri, D.~Kubiznak, C.~Warnick and Y.~Yasui,
  ``Symmetries of the Dirac operator with skew-symmetric torsion,''
  Class.\ Quant.\ Grav.\  {\bf 27} (2010) 185019
  [arXiv:1002.3616 [hep-th]].


  \bibitem{gky1}
  G.~Papadopoulos,
  ``Killing-Yano equations and G-structures,''
  Class.\ Quant.\ Grav.\  {\bf 25} (2008) 105016
  doi:10.1088/0264-9381/25/10/105016
  [arXiv:0712.0542 [hep-th]].


  \bibitem{gky2}
  G.~Papadopoulos,
  ``Killing-Yano Equations with Torsion, Worldline Actions and G-Structures,''
  Class.\ Quant.\ Grav.\  {\bf 29} (2012) 115008
  doi:10.1088/0264-9381/29/11/115008
  [arXiv:1111.6744 [hep-th]].

  \bibitem{santillan}
  O.~P.~Santillan,
  ``Hidden symmetries and supergravity solutions,''
  J.\ Math.\ Phys.\  {\bf 53} (2012) 043509
  doi:10.1063/1.3698087
  [arXiv:1108.0149 [hep-th]].



  \bibitem{gps}
  G.~W.~Gibbons, G.~Papadopoulos and K.~S.~Stelle,
  ``HKT and OKT geometries on soliton black hole moduli spaces,''
  Nucl.\ Phys.\ B {\bf 508} (1997) 623
  doi:10.1016/S0550-3213(97)00599-3
  [hep-th/9706207].

  \bibitem{het1}
  U.~Gran, P.~Lohrmann and G.~Papadopoulos,
  ``The Spinorial geometry of supersymmetric heterotic string backgrounds,''
  JHEP {\bf 0602} (2006) 063
  doi:10.1088/1126-6708/2006/02/063
  [hep-th/0510176].

  \bibitem{het2}
  U.~Gran, G.~Papadopoulos, D.~Roest and P.~Sloane,
  ``Geometry of all supersymmetric type I backgrounds,''
  JHEP {\bf 0708} (2007) 074
  doi:10.1088/1126-6708/2007/08/074
  [hep-th/0703143 [HEP-TH]].

  \bibitem{gutowski}
  J.~B.~Gutowski, D.~Martelli and H.~S.~Reall,
  ``All Supersymmetric solutions of minimal supergravity in six- dimensions,''
  Class.\ Quant.\ Grav.\  {\bf 20} (2003) 5049
  doi:10.1088/0264-9381/20/23/008
  [hep-th/0306235].

  \bibitem{akyol}
  M.~Akyol and G.~Papadopoulos,
  ``Spinorial geometry and Killing spinor equations of 6-D supergravity,''
  Class.\ Quant.\ Grav.\  {\bf 28} (2011) 105001
  doi:10.1088/0264-9381/28/10/105001
  [arXiv:1010.2632 [hep-th]].


  \bibitem{pakis2}
  J.~P.~Gauntlett and S.~Pakis,
  ``The Geometry of D = 11 killing spinors,''
  JHEP {\bf 0304} (2003) 039
  doi:10.1088/1126-6708/2003/04/039
  [hep-th/0212008].

  \bibitem{hull}
  C.~Hull,
  ``Holonomy and symmetry in M theory,''
  hep-th/0305039.


  \bibitem{tsimpis}
  G.~Papadopoulos and D.~Tsimpis,
  ``The Holonomy of IIB supercovariant connection,''
  Class.\ Quant.\ Grav.\  {\bf 20} (2003) L253
  doi:10.1088/0264-9381/20/20/103
  [hep-th/0307127].

  \bibitem{duff}
  M.~J.~Duff and J.~T.~Liu,
  ``Hidden space-time symmetries and generalized holonomy in M theory,''
  Nucl.\ Phys.\ B {\bf 674} (2003) 217
  doi:10.1016/j.nuclphysb.2003.09.019
  [hep-th/0303140].

\end{thebibliography}
\end{document}